\nonstopmode

\documentclass[prb,preprint,showpacs]{revtex4}
\usepackage{graphicx}
\usepackage{color}

\begin{document}
\title{Spin-Transfer Torque Magnetization Reversal in Uniaxial Nanomagnets with Thermal Noise}
\author{D.~Pinna}
\email{daniele.pinna@nyu.edu}
%\affiliation{Department of Physics, New York University, New York, NY 10003, USA}
\author{A.~D.~Kent}
\affiliation{Department of Physics, New York University, New York, NY 10003, USA}
\author{D.~L.~Stein}
\affiliation{Department of Physics, New York University, New York, NY 10003, USA}
\affiliation{Courant Institute of Mathematical Sciences, New York University, New
             York, NY 10012, USA}

\begin{abstract}
We consider the general Landau-Lifshitz-Gilbert (LLG) dynamical theory underlying the magnetization switching rates of a thin film uniaxial magnet subject to spin-torque effects and thermal fluctuations (thermal noise). After discussing the various dynamical regimes governing the switching phenomena, we present analytical results for the mean switching time behavior. Our approach, based on explicitly solving the first passage time problem, allows for a straightforward analysis of the thermally assisted, low spin-torque, switching asymptotics of thin film magnets. To verify our theory, we have developed an efficient GPU-based micromagnetic code to simulate the stochastic LLG dynamics out to millisecond timescales. We explore the effects of geometrical tilts between the spin-current and uniaxial anisotropy axes on the thermally assisted dynamics. We find that even in the absence of axial symmetry, the switching times can be functionally described in a form virtually identical to the collinear case. We further verify that asymptotic behavior is reached fairly slowly, thus quantifying the role of thermal noise in the crossover regime linking deterministic to thermally assisted magnetization reversal. 
%\begin{description}
%\end{description}
\end{abstract}

\pacs{Valid PACS appear here}% PACS, the Physics and Astronomy
\maketitle

%\tableofcontents

\section{Introduction}

More than a decade has passed since spin-torque effects were demonstrated experimentally by the switching of the magnetization of a thin ferromagnetic film when current is passed between it and a pinned ferromagnetic layer~\cite{Slon,Berger,Katine2000,Ozy}. A spin-polarized current passing 
through a small magnetic conductor will deposit spin-angular momentum into the magnetic system. This in turn causes the magnetic 
moment to precess and in some cases even switch direction. This has led to sweeping advances in the field of spintronics through the development and study of spin-valves and magnetic tunnel junctions (see, for example,~\cite{Brataas}). The theoretical approach to such a problem has conventionally been 
to treat the thin ferromagnetic film as a single macrospin in the spirit of Brown~\cite{Brown}. Spin-torque effects are taken 
into account phenomenologically by modifying the macrospin's LLG dynamical equation~\cite{Slon}. A thorough understanding of the phenomena, however, cannot proceed 
without taking into account the effect of thermal fluctuations. This is of particular experimental relevance since spin-transfer effects 
on nanomagnets are often conducted at low currents, where noise is expected to dominate. Recent debate in the literature over the proper exponential
scaling behavior between mean switching time and current shows how the thermally assisted properties of even the simplest magnetic setups leave much to be 
understood~\cite{Koch,Sun, Apalkov, LiZhang, Katine, APL}. The interplay between spin-torque and thermal effects determine the dynamical properties of recent experimental studies on nanopillar devices~\cite{Bedau}. Except at very high currents where the dynamics are predominantly deterministic, the
switching appears to be thermal in nature. Fitting to experimental data requires accurate knowledge of the energetics which, in the realm of spin-torque, 
are hard to come by due to the inherently non-conservative nature of the spin-torque term. 

Theoretical progress has been hindered by the computational power needed to run numerical simulations to the desired degree of accuracy. The LLG equation, modified into its set of coupled stochastic equations can be studied in one of two ways: either by concentrating on the associated Focker-Planck equation or by constructing a stochastic Langevin integrator to be used enough times to gather sufficient statistics on the phenomena~\cite{Palacios}. The latter approach, however, has been unable to extrapolate to long enough times to capture the dynamical extent of the thermal regime. A recent paper by Taniguchi and Imamura~\cite{Taniguchi}, suggests that previous analytics of the thermally assisted dynamics should be revisited. Nonetheless, no numerical simulation has yet been able to evaluate the accuracy of the Taniguchi and Imamura results~\cite{Taniguchi3}. In our paper, we will show that simulations run harnessing the vast computational parallelization capabilities intrinsic in Graphics Processing Unit (GPU) technology for numerical modeling can allow a deeper
probing of such a thermally activated regime.

%--------------------------------------------------------------------------------------------------

\section{General Formalism}

A simple model of a ferromagnet uses a Stoner-Wohlfarth monodomain magnetic body with magnetization $\mathbf{M}$. The body is assumed to have a size $l_m$ along the $\mathbf{e}_y$ direction, and size $a$ in both the $\mathbf{e}_x$ and $\mathbf{e}_z$ directions. The total volume of the object is then $V=a^2l_m$. The shape of the body is assumed to be isotropic and the energy landscape experienced by $\mathbf{M}$ is generally described by three terms: 
an applied field $\mathbf{H}$, 
a uniaxial anisotropy energy $U_K$ with easy axis along $\mathbf{\hat{n}}_K$ in the $\mathbf{e}_x - \mathbf{e}_z$ making an angle $\omega$ with the $\mathbf{e}_z$ axis, and an easy-plane anisotropy $U_p$ in the $\mathbf{e}_y - \mathbf{e}_z$ plane, with normal direction $\mathbf{\hat{n}}_D$ perpendicular to $\mathbf{\hat{n}}_K$. The magnetization $\mathbf{M}$ is assumed to be constant in magnitude and for simplicity normalized into a unit direction vector $\mathbf{m}=\mathbf{M}/|\mathbf{M}|$. A spin-polarized current $J$ enters the magnetic body in the $-\mathbf{e}_y$ direction, with spin polarization factor $\eta$, and spin direction along $\mathbf{e}_z$. The current exits in the same direction, but with its average spin direction aligned to that of $\mathbf{M}$. The self-induced magnetic field of the current is ignored here as the dimension $a$ is considered to be smaller than $1000 \AA$, where the spin-current effects are expected to become dominant over the current induced magnetic field. The standard Landau-Lifshitz-Gilbert (LLG) equation used to describe the dynamics is then written as:

\begin{eqnarray}
\mathbf{\dot{m}}=&-&\gamma'\mathbf{m}\times\mathbf{H}_{\mathrm{eff}}-\alpha\gamma'\mathbf{m}\times\left(\mathbf{m}\times\mathbf{H}_{\mathrm{eff}}\right)\nonumber\\
&-&\gamma' j\mathbf{m}\times\left(\mathbf{m}\times\mathbf{\hat{n}}_p\right)+\gamma'\alpha j\mathbf{m}\times\mathbf{\hat{n}}_p,
\end{eqnarray}
where $\gamma'=\gamma/(1+\alpha^2)$ is the Gilbert ratio, $\gamma$ is the usual gyromagnetic ratio and $j=(\hbar/2e)\eta J$ is the spin-angular momentum deposited per unit time with $\eta = (J_{\uparrow}-J_{\downarrow})/(J_{\uparrow}+J_{\downarrow})$ the spin-polarization factor of incident current $J$. The last two terms describe a vector torque generated by current polarized in the direction $\mathbf{\hat{n}}_p$. These are obtained by assuming that the macrospin absorbs angular momentum from the spin-polarized current only in the direction perpendicular to $\mathbf{m}$.~\cite{Slon}
\newline

To write $\mathbf{H}_{\mathrm{eff}}$ explicitly, we must construct a proper energy landscape for the magnetic body. There are three main components that need to be considered: a uniaxial anisotropy energy $U_K$, an easy-plane anisotropy $U_P$ and a pure external field interaction $U_H$. These are written as follows:

\begin{eqnarray}
&U_K=-K (\mathbf{\hat{n}}_K\cdot\mathbf{m})^2\nonumber\\
&U_P=K_P(\mathbf{\hat{n}}_D\cdot\mathbf{m})^2\nonumber\\
&U_H=-M_S V \mathbf{m}\cdot\mathbf{H}_{ext}\nonumber
\end{eqnarray}
In these equations, $M_S$ is the saturation magnetization, $K_P$ is the easy-plane anisotropy, $K=(1/2) M_S V H_K$ and $H_K$ is the Stoner-Wohlfarth switching field (in units of Teslas). In what follows, we consider the simplified model where we ignore the effects of easy-plane anisotropys and all external magnetic fields are absent. The full energy landscape then becomes $U(\mathbf{m})=U_K + U_H$ and reads:

\begin{equation}
U(\mathbf{m})=-K\left[(\mathbf{\hat{n}}_K\cdot\mathbf{m})^2+2\mathbf{h}\cdot\mathbf{m}\right],
\end{equation}
where $\mathbf{h}=\mathbf{H}_{ext}/H_K$, $\mathbf{\hat{n}}_K$ is the unit vector pointing in the orientation of the uniaxial anisotropy axis, and effects due to external magnetic fields are included for reasons which will be explained in the next section. Such an energy landscape generally selects stable magnetic configurations parallel and anti-parallel to $\mathbf{\hat{n}}_K$. The effective interaction field $\mathbf{H}_{\mathrm{eff}}$ is then given by

\begin{equation}
\mathbf{H}_{\mathrm{eff}}=-\frac{1}{M_S V}\nabla_{\mathbf{m}}U(\mathbf{m})=H_K\left[(\mathbf{\hat{n}}_K\cdot\mathbf{m})\mathbf{\hat{n}}_K+\mathbf{h}\right].
\end{equation}
The symmetries of the problem lead to slightly simplified equations and the deterministic LLG dynamics can then be expressed as:

\begin{eqnarray}
\mathbf{\dot{m}}=&-\mathbf{m}\times\left[(\mathbf{\hat{n}}_K\cdot\mathbf{m})\mathbf{\hat{n}}_K+\mathbf{h}\right]-\alpha\mathbf{m}\times\left[\mathbf{m}\times\left((\mathbf{\hat{n}}_K\cdot\mathbf{m})\mathbf{\hat{n}}_K+\mathbf{h}\right)\right]\nonumber\\
&-\alpha I\mathbf{m}\times\left(\mathbf{m}\times\mathbf{\hat{k}}\right)+\alpha^2 I\mathbf{m}\times\mathbf{\hat{k}},
\end{eqnarray}
where we have defined for future convenience $I=j/(\alpha H_K)$ and introduced the natural timescale $\tau=\gamma' H_K t$.

%---------------------------------------------------------------------------------------------------

\section{Thermal Effects}

Thermal effects are included by considering uncorrelated fluctuations in the effective interaction field: $\mathbf{H}_{\mathrm{eff}}\rightarrow\mathbf{H}_{\mathrm{eff}}+\mathbf{H}_{th}$. These transform the LLG equation into its Langevin form. We model the stochastic contribution $\mathbf{H}_{th}$ by specifying its correlation properties, namely:

\begin{eqnarray}
&\langle \mathbf{H}_{th}\rangle=0 \nonumber\\
&\langle H_{th,i}(t)H_{th,k}(t')\rangle=2D\delta_{i,k}\delta(t-t')
\end{eqnarray} 
The effect of the random torque $\mathbf{H}_{th}$ is to produce a diffusive random walk on the surface of the $\mathbf{M}$-sphere. An associated Focker-Planck equation describing such dynamics was constructed by Brown~\cite{Brown}. At long times, the system attains thermal equilibrium and, by the fluctuation-dissipation theorem we have:

\begin{equation}
D=\frac{\alpha k_BT}{2K(1+\alpha^2)}=\frac{\alpha}{2(1+\alpha^2)\xi}.
\end{equation}
It is convenient to introduce the notation $K/k_BT = \xi$, representing the natural barrier height of the uniaxial anisotropy energy. 

%\newline
Setting now the external magnetic field to zero and considering only thermal fluctuations, the stochastic LLG equation reads:

\begin{equation}
\dot{m}_i=A_i(\mathbf{m})+B_{ik}(\mathbf{m})\circ H_{th,k}
\end{equation}
where

\begin{eqnarray}
\mathbf{A}(\mathbf{m})&=&\alpha I\left[\alpha\mathbf{m}\times\mathbf{\hat{k}}-\mathbf{m}\times\left(\mathbf{m}\times\mathbf{\hat{k}}\right)\right]\nonumber\\
&-&(\mathbf{\hat{n}}_K\cdot\mathbf{m})\left[\mathbf{m}\times\mathbf{\hat{n}}_K-\alpha\left(\mathbf{\hat{n}}_K-(\mathbf{\hat{n}}_K\cdot\mathbf{m})\mathbf{m}\right)\right],\nonumber\\
B_{ik}(\mathbf{m})&=&-\epsilon_{ijk}m_j-\alpha(m_i m_k - \delta_{ik}).\nonumber
\end{eqnarray}
and `$\circ H_{th,k}$' means to interpret our stochastic dynamics in the sense of Stratonovich\footnote{In modeling thermal effects through stochastic contributions in a system like ours, we are interested in considering the white noise limit of a potentially colred noise process. Stratonovich calculus is then to be preferred over \^Ito calculus as ascertained by the Wong-Zakai theorem.} calculus in treating the multiplicative noise terms~\cite{Karatsas}.

We numerically solve the above Langevin equations by using a standard second order Heun scheme to ensure proper convergence to the Stratonovich calculus. At each time step, the strength of the random kicks is given by the fluctuation-dissipation theorem. Statistics were gathered from an ensemble of $5000$ events with a natural integration stepsize of $0.01$. For concreteness, we set the Landau damping constant $\alpha=0.04$. Different barrier heights were explored although the main results in this paper are shown for a barrier height of $\xi=80$. To explore the simulations for long time regimes, the necessary events were simulated in parallel on an NVidia Tesla C2050 graphics card. To generate the large number of necessary random numbers, we chose a proven combination~\cite{Nguyen} of the three-component combined Tausworthe ``taus88"~\cite{Ecuyer} and the 32-bit ``Quick and Dirty" LCG~\cite{Press}. The hybrid generator provides an overall period of around $2^{121}$. 

%----------------------------------------------------------------------------------

\section{Switching Dynamics}

In experiments, one is generally interested in understanding how the interplay between thermal noise and spin torque effects switch an initial magnetic orientation from parallel to anti-parallel and vice versa. The role of spin-torque can be clarified by considering how energy is pumped in the system, from an energy landscape point of view. As in the previous section, the magnetic energy of the monodomain is:

\begin{equation}
U(\mathbf{m})=-K(\mathbf{\hat{n}}_K\cdot\mathbf{m})^2.\nonumber
\end{equation}
The change in energy over time can be obtained after some straightforward algebra and is found to be:

\begin{equation}
\frac{1}{M_sVH_K}\dot{U}=-\left[\alpha\mathbf{m}\times\mathbf{H}_{\mathrm{eff}}-I(\alpha\mathbf{\hat{k}}-\mathbf{m}\times\mathbf{\hat{k}})\right]\cdot\left(\mathbf{m}\times\mathbf{H}_{\mathrm{eff}}\right).
\end{equation}
This expression shows how current pumps energy into the system. In the absence of current, the damping dissipates energy and, as one would expect, the dynamical flow is toward the minimum energy configuration. The sign preceeding the current term allows the expression to become positive in certain regions of magnetic configuration space.
Furthermore, by averaging over constant energy trajectories, one can construct an equivalent dynamical flow equation in energy space. This kind of approach has already been used in the literature~\cite{Apalkov} and can lead to the appearance of stable limit cycles at currents less than the critical current as can be intuitively inferred by considering which constant energy trajectories lead to a canceling of the flow in (11).

Starting from an initially stable magnetic state, spin-torque effects will tend to drive the magnetization toward the current's polarization axis. Once the current is turned off, the projection of the magnetization vector along the uniaxial anisotropy axis will almost surely determine which stable energy state (parallel or anti-parallel) the magnetic system will relax to as long as the energy barrier $\xi$ is large enough. As such, switching dynamics are best studied by projecting equations (8) along the uniaxial anisotropy axis $\mathbf{\hat{n}}_K$. One then obtains a stochastic differential equation describing the dynamics of such a projection:

\begin{eqnarray}
\dot{q}&=&\alpha\left[(n_zI+q)(1-q^2)+n_zIq(q-\frac{m_z}{n_z})\right]\nonumber\\
&+&\alpha^2In_xm_y+\sqrt{\frac{\alpha}{\xi}(1-q^2)}\circ\dot{W}.
\end{eqnarray}
In the above equation, we have defined $q \equiv \mathbf{m}\cdot\mathbf{\hat{n}}_K$; $n_z$ and $n_x$ are the projections of the uniaxial anisotropy axis respectively on the $z$ and $x$ axes. Furthermore, the multiple stochastic contributions are assumed to be gaussian random variables with identical average and space dependent variance. $\dot{W}$ is a standard mean zero, variance 1, Wiener process, and its prefactor explicitly expresses the strength of the stochastic contribution~\cite{Karatsas}. As will be shown in the following subsections, (12) is a convenient analytical tool in specific scenarios. In general, though, it is not useful as it explicitly depends on the dynamics of both the $m_z$ and $m_y$ components of the magnetization.

Numerically, we can solve (8) directly. In all our simulations, the initial ensemble of magnetizations was taken to be Boltzmann-distributed along the anti-parallel orientation. We assume that the energy barrier height is so large that, before current effects are activated, thermalization has only been achieved within the antiparallel energy well and no states have had time to thermally switch to the parallel orientation on their own. A typical histogram of magnetic orientations at a given time is shown in Figure 1.

\begin{figure}
  \centerline{\includegraphics [clip,width=4.0in, angle=0]{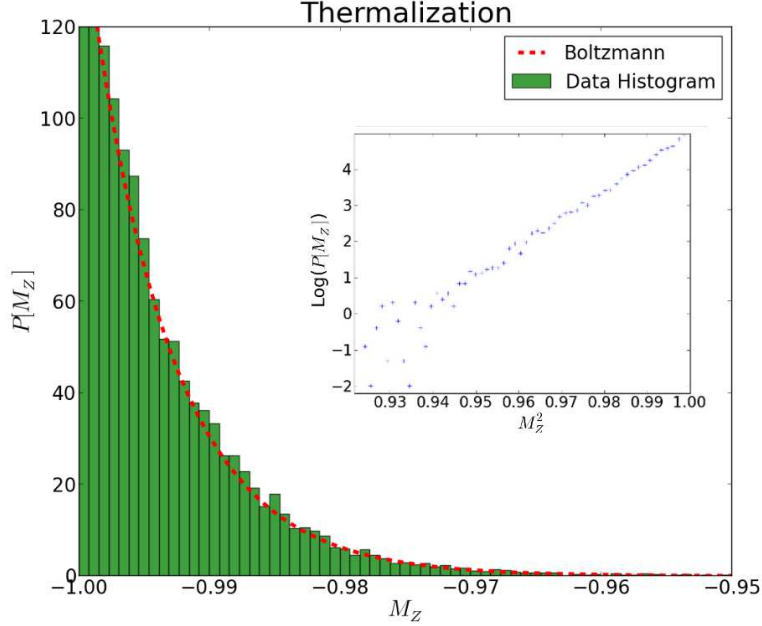}}
  \caption{Histogram distribution of $m_z$ after letting the magnetic system relax to thermal equilibrium ($10^3$ natural time units). The overlayed red dashed line is the theoretical equilibrium Boltzmann distribution. In the inset we show a semilog-plot of the probability vs. $m_z^2$ dependency. As expected, the data scales linearly.}
  \label{F1}
\end{figure}

We now turn on a current and allow the system to evolve for a fixed amount of time. Once this time has passed, we use the projection rule expressed above to evaluate what fraction of the ensemble has effectively switched from the anti-parallel to the parallel state.
 
%----------------------------------------------------------------------------------

\section{Collinear Spin-Torque Model}

Having derived the necessary expressions for our macrospin model dynamics, it is useful to consider the following simplification. Let us take the uniaxial anisotropy and spin-current axes to be collinear, namely, $\mathbf{\hat{n}}_K \equiv \mathbf{\hat{n}}_p \equiv \mathbf{\hat{k}}$. In such a scenario, the stochastic LLG equation simplifies significantly. In particular, (11) reduces to the simplified form:

\begin{equation}
\dot{q}=\alpha(I+q)(1-q^2)+\sqrt{\frac{\alpha}{\xi}(1-q^2)}\circ\dot{W}.
\end{equation}
In this symmetric scenario, $q$ coincides with $m_z$ and magnetization reversal has been reduced to a straightforward 1-D problem. For a general value of $I<1$, the evolution of $q$ has two local minima and a saddle. The two stable configurations are at $q=-1$ and $q=1$, while the saddle is located at $q=-I$. For currents $I>I_c=1$ there is only one stable minimum. Above critical current, spin torque pushes all magnetic configurations toward the parallel $q=1$ state.

\subsection{Collinear High Current Regime}
In the high current regime $I\gg I_c$ we expect the deterministic dynamics to dominate over thermal effects. We refer to this also as ballistic evolution interchangeably. The determistic contribution of (12) can then be solved analytically given an initial configuration $q \equiv m_z = -m_0$. The switching time $\tau_s$ will simply be the time taken to get from some $m_z=-m_0<0$ to $m_z=0$ and reads:

\begin{eqnarray}
\tau_s(m_0) &=& \frac{1}{\alpha}\int_{-m_0}^0\frac{dm}{(I+m)(1-m^2)} \nonumber\\
&=& \frac{1}{2\alpha(I^2-1)}\bigg\{I\log\left[\frac{1+m_0}{1-m_0}\right]\nonumber\\
&&-\log\left[1-m_0^2\right]-2\log\left[\frac{I}{I-m_0}\right]\bigg\}.	
\end{eqnarray} 
Since the magnetic states are considered to be in thermal equilibrium before the current is turned on, one should average the above result over the equilibrium Boltzmann distribution in the starting well to obtain the average switching time $\langle \tau_s \rangle_B$. For $\xi$ large enough, such an initial distribution will be:

\begin{equation}
\rho_B(m) = \frac{\sqrt{\xi}\exp[-\xi]}{F[\sqrt{\xi}]}\exp[\xi m^2],
\end{equation}
where $F[x]=\exp(-x^2)\int_0^x\exp(y^2)dy$ is Dawson's integral. This expression can be used to compute the average switching time numerically.

As the intensity of spin-currents becomes closer to $I_c$, thermal effects increasingly contribute. Moreover, diffusion gradients add to the deterministic drift, which can be shown explicitly by writing (13) in its equivalent \^Ito form. Doing so leads us to a first correction of the ballistic dynamics due to thermal influences. The z-component behavior then reads:

\begin{equation}
\dot{m}_z = \alpha(I+m_z)(1-m_z^2)-\frac{\alpha}{2\xi}m_z+\sqrt{\frac{\alpha}{\xi}(1-m_z^2)}\dot{W}
\end{equation}
The first term on the right hand side is still the ballistic flow that we have just discussed. The second term is the desired diffusion-gradient drift term. The contribution of such a term generates a net motion away from the stable minima of the ballistic equations as one expects to see under the influence of thermal effects. Again, we can solve the drift dominated flow analytically to compute the switching time. Considering diffusion-gradient drift, this reads:

\begin{eqnarray}
\tau_s(m_0) &=& \frac{1}{\alpha}\int_{-m_0}^0\frac{dm}{(I+m)(1-m^2)-(m/2\xi)}\nonumber\\
 &=& \frac{1}{\alpha}\sum_j \frac{\log\left[(w_j-m_0)/w_j\right]}{3 w_j^2-2 I w_j-(1-\frac{1}{2\xi})}	
\end{eqnarray} 
Where the $w_j$ are the three zeros of the cubic equation $ w^3-I w^2-(1-\frac{1}{2\xi})w+I = 0$. As before, the average switching $\langle \tau_s\rangle_B$ time will simply be given by averaging numerically over the Boltzmann distribution $\rho_B$. In Figure 2, the reader can see how well these two limiting results fit the simulation data. As expected both expressions coincide in the limit of high currents.

\begin{figure}
  \centerline{\includegraphics [clip,width=4.0in, angle=0]{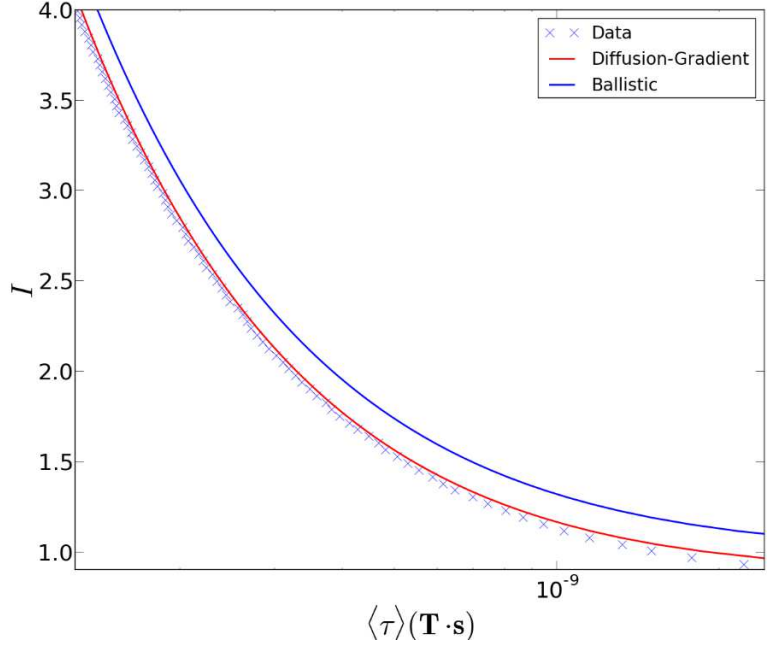}}
  \caption{Blue line shows the fit of the ballistic limit to the numerical data (in blue crosses). Red line shows the improvement obtained by including diffusion gradient terms. Times are shown in units of ($s\cdot T$) where $T$ stands for Tesla: real time is obtained upon division by $H_K$.}
  \label{F2}
\end{figure}

%---------------------------------------------------------------------------------------------------------------------------------------

\section{Uniaxial Tilt}

In the high current regime ($I\gg I_c$), where $\mathbf{\hat{n}}_K=\mathbf{\hat{k}}$ (i.e. the uniaxial tilt is aligned with the z-axis), the ballistic equation for $m_z$ was shown to decouple from the other components, and the dynamics became one dimensional and deterministic. For the more general case where the uniaxial anisotropy axis may have any tilt with respect to the z-axis, such a critical current is not as intuitively defined. Unlike the collinear limit, a critical current, above which all magnetic states perceive a net flow towards an increasing projection, does not exist. One can plot $\dot{q}\equiv\mathbf{\dot{m}}\cdot\mathbf{\hat{n}}_K$ over the unit sphere to see what regions allow for an increasing and decreasing projection as the current is changed. An example of this is shown in Figure 3.

\begin{figure}
  \centerline{\includegraphics [clip,width=4.0in, angle=0]{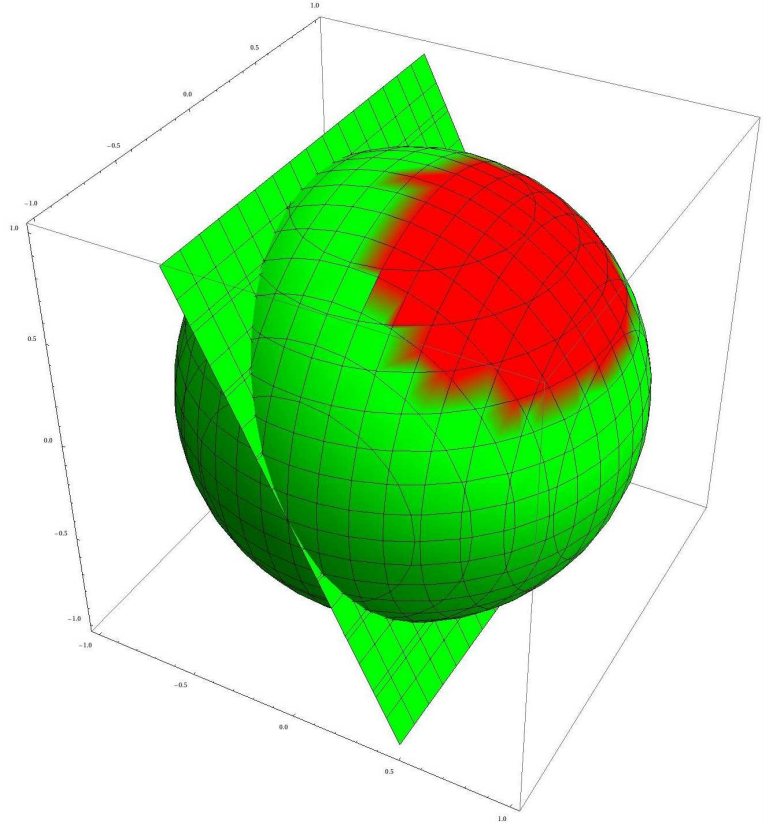}}
  \caption{$\dot{q}$: green $>0$, red $<0$ for applied current $I=5$. The plane dissecting the sphere is perpendicular to the uniaxial anistropy axis. Its intersection with the sphere selects the regions with highest uniaxial anisotropy energy.}
  \label{F3}
\end{figure}

Unfortunately, regions characterizing negative projection flow can be shown to persist at all currents. The approach is refined by requiring that on average, over constant-energy precessional trajectories, the flow is toward the positive uniaxial anisotropy axis: $\langle \mathbf{\dot{m}}\cdot\mathbf{\hat{n}}_K\rangle > 0$. Such trajectories are found by solving the flow equations with $h_s=\alpha=0$. Solutions correspond to circular libations about the uniaxial anisotropy axis. The critical current is then redefined to be the minimum current at which the average projectional flow is positive at all possible precessional energies. This is easily done and results in:

\begin{equation}
I \geq max_{\epsilon} \left[\frac{-\epsilon}{\cos{\omega}}\right] = \frac{1}{\cos(\omega)} = I_{crit},
\end{equation}
thus allowing for a direct comparison of dynamical switching results between different angular configurations of uniaxial tilt. In our discussion of (12) we mentioned how in the general case, presenting uniaxial tilt, there is no way to reduce the dimensionality of the full dynamical equations. In fact, in the presence of tilt, precessional trajectories might allow for a magnetization state to temporarily transit through a $q>0$, ``switched" configuration, even though it might spend the majority of its orbit in a $q<0$, ``unswitched" configuration. This allows for a much richer mean switching time behavior, especially above critical current, as shown in Figure 4 and discussed more in depth later.

\begin{figure}
  \centerline{\includegraphics [clip,width=4.0in, angle=0]{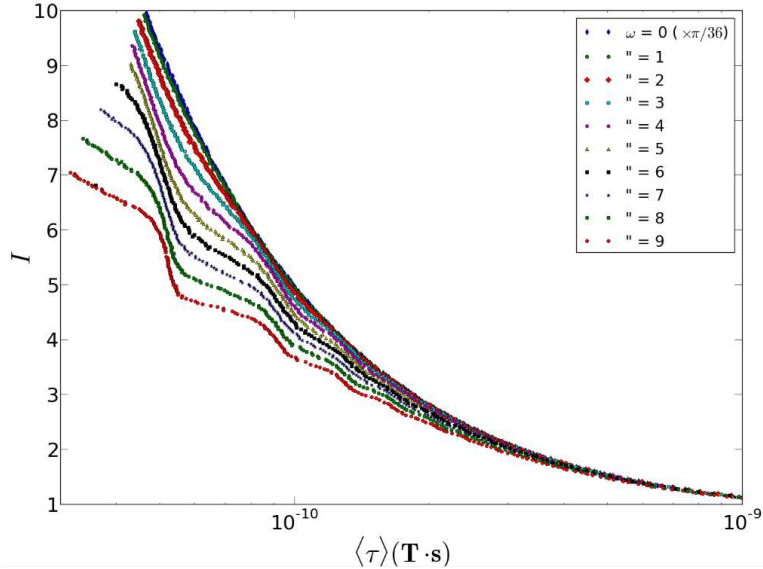}}
  \caption{Mean switching time behavior for various angular tilts above critical current. Each set of data is rescaled by its critical current such that all data plotted has $I_c = 1$. Angular tilts are shown in the legend in units of $\pi/36$ such that the smallest angular tilt is $0$ and the largest is $\pi/4$. Times are shown in units of ($s\cdot T$) where $T$ stands for Tesla: real time is obtained upon division by $H_K$.}
  \label{F4}
\end{figure}

%---------------------------------------------------------------------------------------------------------------------------------------

\section{Thermally Activated Regime}

For currents $I<I_c$ the switching relies on thermal effects to stochastically push the magnetization from one energy minimum to the other. It is of interest to understand how switching probabilities and switching times depend on temperature and applied current. This is easily done in stochastic systems with gradient flow. In such cases an energy landscape exists and a steady state probability distribution can be constructed to be used via Kramer's theory in deriving approximate low-noise switcing probabilities. 

Unfortunately, though, spin-torque effects introduce a non-gradient term, and the resulting LLG equation does not admit an energy landscape in the presence of applied current. The collinear simplification, however, is an exception. As already described, in the absence of uniaxial tilt the dynamics become effectively one dimensional since the $m_z$ component decouples from the other magnetization components. Consider then (13): because it is decoupled from the other degrees of freedom, we can construct a corresponding one-component Focker-Planck equation. The evolution in time of the distribution of $q$ is then:

\begin{equation}
\partial_t \rho(q,t) = \hat{L}[\rho](q,t),\nonumber
\end{equation}
where

\begin{equation}
\hat{L}[f]=-\alpha\partial_q \left[(q+I)(1-q^2)-\frac{1}{2\xi}(1-q^2)\partial_q\right]f.\nonumber
\end{equation}

For high energy barriers and low currents, the switching events from one basin to the other are expected to be rare. The probability of a double reversal should be even smaller. We therefore model the magnetization reversal as a mean first passage time (MFPT) problem with absorbing boundaries at the saddle point. The MFPT will then be given by~\cite{Zwanzig} the solution of the adjoint equation ($\hat{L}^{\dagger}\langle\tau\rangle(q)=-1$):

\begin{equation}
\frac{\alpha}{2\xi}\exp(-\xi(q+I)^2)\partial_q\left[(1-q^2)\exp(\xi(q+I)^2)\right]\partial_q\tau(q)=-1\nonumber
\end{equation}
subject to the boundary condition $\langle\tau\rangle(0)=0$. This can be solved to give:

\begin{equation}
\langle\tau\rangle(q)=\frac{2\xi}{\alpha}\int_q^{-I} du \frac{\exp(-\xi(u+I)^2)}{1-u^2} \int_{-1}^u ds \exp(\xi(s+I)^2).\nonumber
\end{equation}
The rightmost integral can be computed explicitely. Retaining only dominant terms, the final integral can be computed by saddlepoint approximation to give:

\begin{equation}
\langle\tau\rangle\simeq\frac{\sqrt{\pi}}{\alpha}\frac{\exp(\xi(1-I)^2)F(\sqrt{\xi}(1-I))}{1-I^2}.
\end{equation}
Such a square exponential dependence has recently been derived by Taniguchi and Imamura~\cite{Taniguchi,Taniguchi2}, although a $\tau\propto\exp(\xi(1-I))$ dependence, proposed elsewhere in the literature~\cite{Koch, Apalkov, LiZhang}, has also been successfully used to fit experimental data~\cite{Bedau}. 

To decide between these experimental dependences, we fit the scaling behaviors in Figure 5, along with the theoretical prediction from (24). The square exponential dependence fits the data better, confirming our analytical results. Furthermore, comparison to the full theoretical prediction demonstrates that even for mean switching times of the order $10^{-1}$ milliseconds, asymptoticity still is not fully achieved.

\begin{figure}
	\begin{center}
	\centerline{\includegraphics [clip,width=4.0in, angle=0]{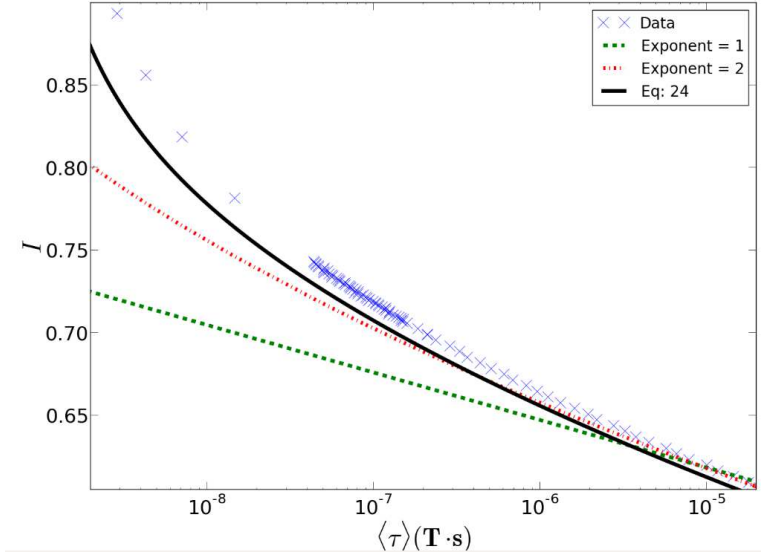}}
	\caption{Mean switching time behavior in the sub-critical low current regime. Times are shown in units of ($s\cdot T$) where $T$ stands for Tesla: real time is obtained upon division by $H_K$. The red and green line are born by fitting to the data the functional form $\langle\tau\rangle=C\exp(-\xi(1-I)^{\mu})$, where $\mu$ is the debated exponent (either $1$ or $2$) and $C$ is deduced numerically. The red curve fits the numerical data asymptotically better the green curve.}
	\label{F5}
	\end{center}
\end{figure}

All that remains is to consider the effects of angular tilt on the switching properties in the thermally activated regime. Insight into this problem can be obtained by invoking (12) again. For small values of $\alpha$, the term in square brackets is of leading order over the second ballistic term depending on $m_y$. This allows us, in the small $\alpha$ regime, to neglect the second ballistic term altogether. 

We now concentrate on the behavior of the term in square brackets. For low sub-critial currents, switching will depend on thermal activation for the most part. We expect an initially anti-parallel configuration to not diffuse very far away from its local energy minima. We expect it to remain that way until a strong enough thermal kick manages to drive it across the energy barrier. Because of this, the second term appearing in the square brackets will generally be close to zero as the particle awaits thermal switching. To make the statement more precise, one can imagine the magnetic state precessing many times before actually making it over the saddle. The second term can then be averaged over a constant energy trajectory and shown to vanish identically. Hence, in the subcritical regime, (12) can be rewritten in the following approximate form:

\begin{equation}
\dot{q}=\alpha(n_zI+q)(1-q^2)+\sqrt{\frac{\alpha}{\xi}(1-q^2)}\circ\dot{W}.
\end{equation}
This, is reminiscent of the 1D projectional dynamics discussed in relation to the collinear limit, and shown explicitly in (13). The only difference between the two is the substitution $I \rightarrow n_zI$; recall because $n_z=cos(\omega)$, $n_zI=I/I_c$. In other words, the thermally activated dynamics are the same for all angular separations up to a rescaling by the critical current. We then expect that the mean switching time dependences remain functionally identical to the collinear case for all uniaxial tilts. We have confirmed this by comparison with data from our simulations, and the results are shown in Figure 6. As predicted, all mean switching time data from different uniaxial tilts collapses on top of each other after a rescaling by each tilt's proper critical current. 

In comparing our scaling relationships between current and mean switching time with the previous literature, a subtle issue must be addressed. Results obtained by Apalkov and Visscher~\cite{Apalkov} rely upon an initial averaging of the dynamics in energy space over constant energy trajectories (limit for small damping) and only subsequently applying weak noise methods to extrapolate switching time dependences. The small damping and weak noise limits are singular and the order in which they are taken is important. Both limits radically alter the form of the equations: whereas both limits suppress thermal effects, the first also severely restrains the deterministic evolution of the magnetic system. Our approach considers the weak noise limit and, only in discussing the effects of an angular tilt between polarization and easy axes do we employ the small damping averaging technique to obtain functional forms for the mean switching time. The switching time data shown seems to justify, in this particular, an interchangeability between these two limits. More generally, however, one should not expect the two limits to commute.

\begin{figure}
	\begin{center}
	\centerline{\includegraphics [clip,width=4.0in, angle=0]{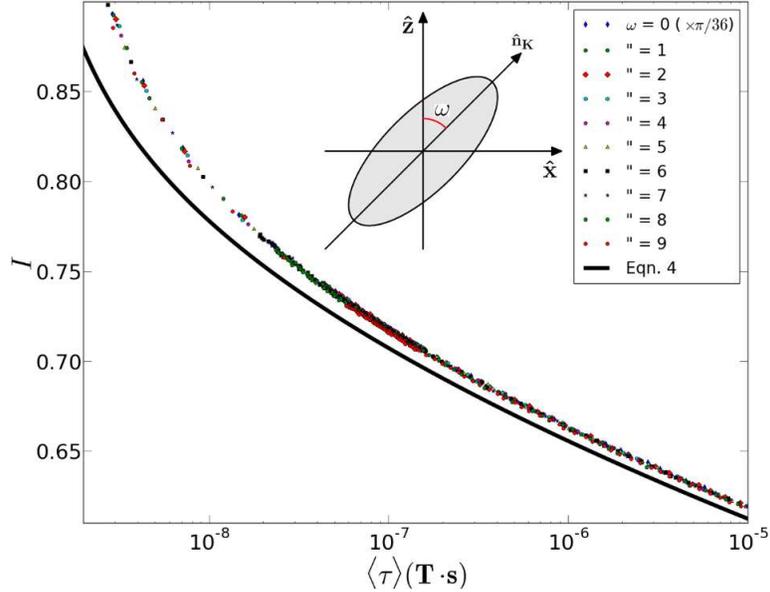}}
	\caption{Mean switching time behavior in the sub-critical low current regime. Various uniaxial tilts are compared by rescaling all data by the appropriate critical current value. Times are shown in units of ($s\cdot T$) where $T$ stands for Tesla: real time is obtained upon division by $H_K$.}
	\label{F6}
	\end{center}
\end{figure}

%-------------------------------------------------------------------------------------

\section{Switching Time Probability Curves}

Up until now, we have analyzed the main properties of spin-torque induced switching dynamics by concentrating solely on the mean switching times.
In experiments, though, one generally constructs full probability curves. The probability that a given magnetic particle has a switching time $\tau_s\leq\tau$ can be explicitely written as:

\begin{eqnarray}
P[\tau_s\leq\tau]&=&\int_0^{m(\tau)}dx\rho_B(x)\nonumber\\
&=&\exp[-\xi(1-m(\tau)^2)]\frac{F[\sqrt{\xi}m(\tau)]}{F[\sqrt{\xi}]},
\end{eqnarray}
where $m(\tau)$ is the initial magnitization that is switched deterministically in time $\tau$.
Once one has evaluated $m(\tau)$, the probability curve follows. Ideally, in the ballistic regime, one would like to invert the ballistic equations. Unfortunately, though, the solutions of such ballistic equations are generally transcendental and cannot be inverted analytically. Even in the simpler collinear case, as can be seen from equations (14) and (18), no analytical inversion is possible. One must instead compute the inversion numerically\footnote{Easily achieved thanks to the monotonicity of their form.}. Nonetheless, one can construct appropriate analytical approximations by inverting the dominant terms in the expressions. In the case of (14), for example, one has that for currents much larger than the critical current:

\begin{equation}
\tau(m)\simeq\frac{I}{2\alpha(I^2-1)}\log[\frac{1+m}{1-m}]
\end{equation}
which can be inverted to give:

\begin{equation}
m(\tau)=\tanh[\alpha\tau\frac{I^2-1}{I}].
\end{equation}
Plugging into expression (25) for the $\tau$ probability curve, one has:

\begin{equation}
P[\tau_s\leq\tau]=\exp\left[-\frac{\xi}{\cosh^2[\alpha\tau\frac{I^2-1}{I}]}\right]\frac{F[\sqrt{\xi}\tanh[\alpha\tau\frac{I^2-1}{I}]]}{F[\sqrt{\xi}]},
\end{equation}
This expression can be truncated to a simpler form by noting that the leading exponential term dominates over the ratio of Dawson functions. Furthermore, if one considers the limit of large values for $\tau$ (or, analogously, $I\gg 1$), the `$\cosh$' can be also approximated by its leading exponential term. We are finally left with:

\begin{eqnarray}
P[\tau_s\leq\tau]&\simeq&\exp\left[-\frac{\xi}{\cosh^2[\alpha\tau\frac{I^2-1}{I}]}\right]\nonumber\\
&\sim&\exp\left[-4\xi\exp[-2\alpha\tau\frac{I^2-1}{I}]\right],
\end{eqnarray}
which is very similar in form to what has already been reported and used for fitting in the literature~\cite{Sun, Bedau}.\newline

In the low current regime, one constructs probability curves by considering the mean switching time and modeling a purely thermal reversal as a decay process with rate given by equation (24). The fraction of switched states then vary in time as:

\begin{equation}
P(m_z>0) = 1-\exp(-t/\langle\tau\rangle).
\end{equation}
Upon introducing uniaxial tilt, precessional effects can be witnessed directly on the switching probability curves in the super-critical regime. One expects that in the initial phases of switching, the fraction of switched states is sensitive to the time at which the current is turned off. One may accidentally turn off the current during a moment of transient passage through the switched region along the precessional orbit. This was checked and verified from our numerical simulations (see Figure 7). More generally, effects similar to the ``waviness" seen in the mean switching time curves can be seen in the probability curves as well as the angle of uniaxial tilt is allowed to vary (see Figure 8). Only a numerical solution of the LLG equation can bring such subtleties to light. 

\begin{figure}
	\begin{center}
	\centerline{\includegraphics [clip,width=4.0in, angle=0]{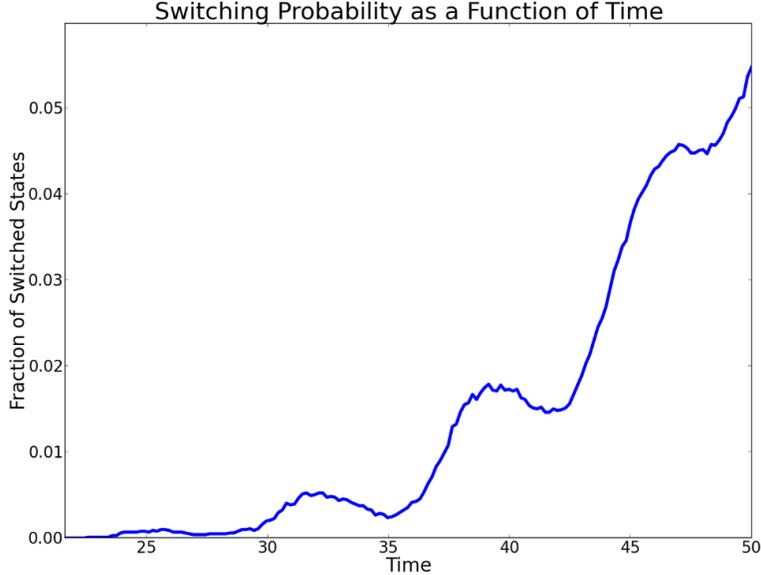}}
	\caption{Influence of precessional orbits on transient switching as seen from the switching time probability curve in the supercritical current regime. The case shown is that of an angular tilt of $\pi/4$ subject to a current intensity of $1.5$ times the critical current. The non-monotonicity in the probability curve shows the existence of transient switching.}
	\label{F7}
	\end{center}
\end{figure}

\begin{figure}
	\begin{center}
	\centerline{\includegraphics [clip,width=4.0in, angle=0]{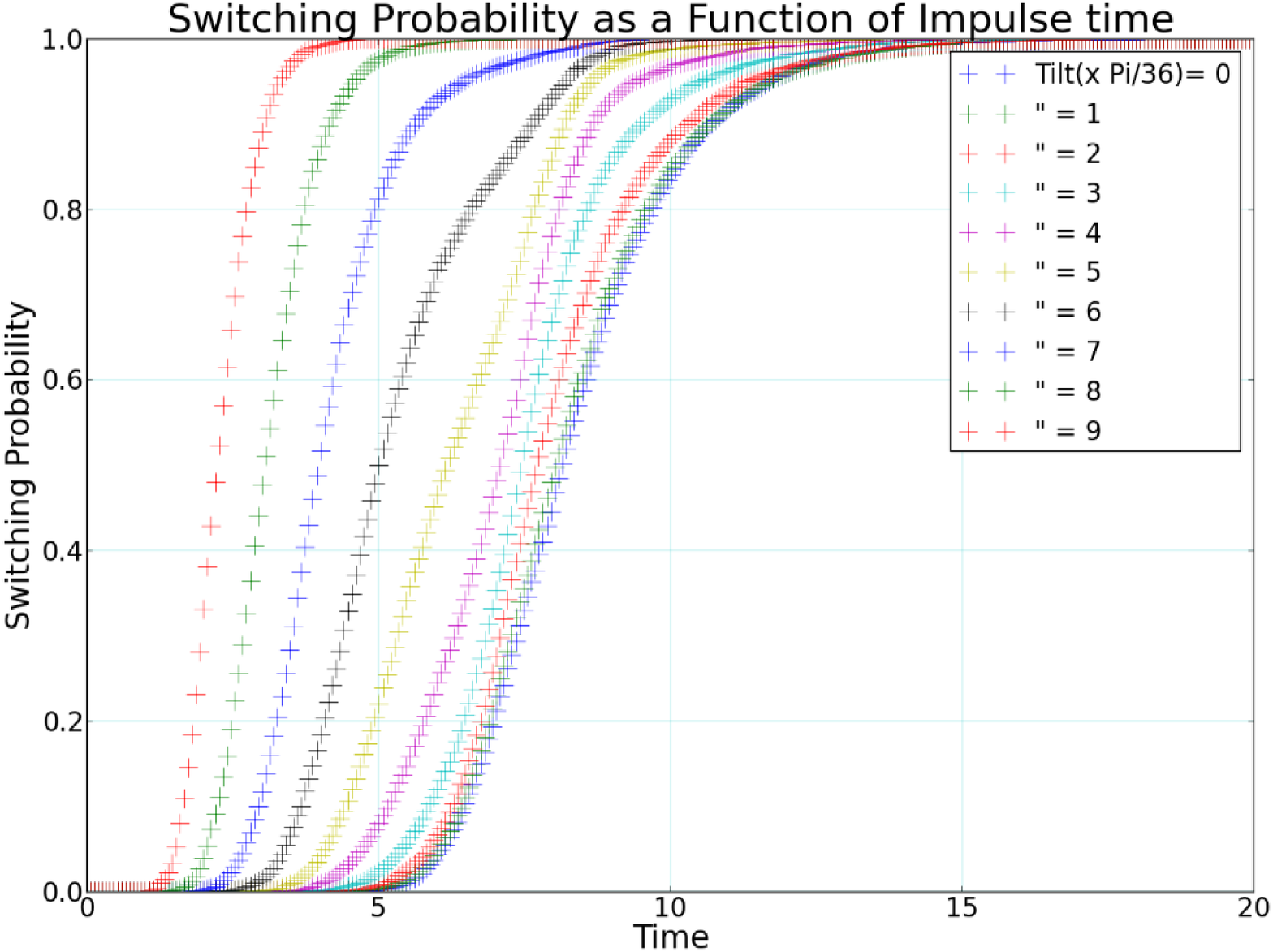}}
	\caption{Spin-torque induced switching time probability curves for various angular configurations of uniaxial tilt (a sample normalized current of $10$ was used).}
	\label{F8}
	\end{center}
\end{figure}

%-----------------------------------------------
\section{Conclusion}

We have constructed the theory underlying the dynamics of a uniaxial macrospin subject to both thermal fluctuation 
and spin-torque effects. We then studied the subtle interplay between these two effects in aiding magnetization reversal
between energy minima in a magnetically bistable system. Two regimes stand out in such a theory: a ballistic regime dominated 
by the deterministic flow and a thermally activated regime where reversal is dominated by noise. In the ballistic regime we discussed
how to approximate the mean switching time behavior and found that corrections due to the diffusion-gradient term, arising from the stochastic equations, allow one to model the dynamics more accurately.

In the thermally activated regime, we solved the relevant mean first passage time problem and obtained an expression for the
dependence of mean switching time on applied current. In doing so, the correct scaling was shown to be $\langle\tau\rangle\propto\exp(-\xi(1-I)^2)$,
in contrast to the prevailing view that $\langle\tau\rangle\propto\exp(-\xi(1-I))$. Analytical results were then compared 
with detailed numerical simulations of the stochastic LLG equation. The massively parallel capabilities of our GPU devices has allowed us to explore 
the behavior of macrospin dynamics over six orders of temporal magnitude. Comparing to our analytical results, we suggest that the thermal asymptotic behavior is achieved very slowly with respect to the switching timescales that have been probed experimentally.

Different geometrical configurations of the uniaxial anisotropy axis with respect to the spin-current axis were shown to influence
the thermally activated regime very minimally inasmuch as the currents were rescaled by the proper critical current of the angular setup.
Only in the super-critical regime were distinctions shown to exist due to complex precessional and transient switching behavior.

These results have important implications for the analysis of experimental data in which measurements of the switching time versus current pulse amplitude are used to determine the energy barrier to magnetization reversal. Clearly use of the correct asymptotic scaling form is essential to properly determine the energy barrier to reversal. The energy barrier, in turn, is very important in assessing the thermal stability of magnetic states of thin film elements that are being developed for long term data storage in STT-MRAM. Further work should address how these results extend to systems with easy plane anisotropy and situations in which the nanomagnet has internal degrees of freedom, leading to a break down of the macrospin approximation.

We also note that current flow is a source of shot noise, which at low frequencies
acts like a white-noise source in much the same way as thermal noise. 
It is therefore interesting to understand when this additional source of noise plays a role.
For a magnetic layer coupled to unpolarized leads, the current induced noise on the magnetization dynamics was found to be 
$\frac{\Gamma_L/\Gamma_R}{(1+\Gamma_L/\Gamma_R)^2}V$~\cite{Mitra06}, where $V$ is the voltage drop across the magnetic layer, while $\Gamma_L/\Gamma_R$ is a dimensionless ratio characterizing the coupling strength of the magnetic layer to the left (L)
and right (R) leads. Thus the noise is maximal ($V/4$) for perfectly symmetrical couplings, and is smaller
in the limit of highly asymmetric contacts. This basic behavior, and the
order of magnitude of the effect, is not likely to be modified by polarized leads. 
We argue that the temperatures at which the experiments have been performed
current noise effects are not important.  The experiments are performed at room temperature where $T=300$ K. For an all-metallic
device, such as a spin-valve nanopillar, the couplings are nearly symmetrical and at the critical current a typical voltage drop across the magnetic layer is less than $10\; \mu$V or, equivalently, $1$ K. For a magnetic tunnel junction device $V$ can be $\sim 1$ V. However, in this case the coupling is asymmetric. One lead (L) forms a magnetic tunnel junction with the nanomagnet, while the other (R) a metallic contact. This gives $\Gamma_R/\Gamma_L > 10^4$ and a relevant energy $\sim 1$ K, again far lower than room temperature.
It appears that current induced noise can only be important at room temperature for a nanomagnet coupled symmetrically between two tunnel barriers.

%-----------------------------------------------------------------------

\begin{acknowledgments}
The authors would like to acknowledge A. MacFadyen, Aditi Mitra and J. Z. Sun for many useful discussions and comments leading to this paper. This research was supported by NSF-DMR-100657 and PHY0965015.

\end{acknowledgments}


\begin{thebibliography}{99}

\bibitem{Slon} J. Slonczewski, J. Magn. Magn. Mater. {\bf159}, L1 (1996).
\bibitem{Berger} L. Berger, Physical Review B {\bf 54}, 9353 (1996).
\bibitem{Katine2000} J. A. Katine, F. J. Albert, and R. A. Buhrman, Phys. Rev. Lett. {\bf 84}, 3149–3152 (2000)
\bibitem{Ozy} B. Özyilmaz, A. D. Kent, J. Z. Sun, M. J. Rooks, and R. H. Koch, Phys. Rev. Lett. {\bf 93}, 176604 (2004)
\bibitem{Brown} W. F. Brown, Phys. Rev. {\bf135}, 5 (1963).
\bibitem{Brataas} A. Brataas, A. D. Kent, H. Ohno, Nature Materials {\bf 11}, 372 (2012).
\bibitem{Koch} R. H. Koch, J. A. Katine, and J. Z. Sun Phys. Rev. Lett {\bf 92} no. 8, (2004) .
\bibitem{Sun} J. Z. Sun, Phys. Rev. B {\bf62}, 1 (2000).
\bibitem{Apalkov} D. M. Apalkov and P. B. Visscher, Phys. Rev. B {\bf72}, 180405R (2005).
\bibitem{LiZhang} Z. Li and S. Zhang, Phys. Rev. B {\bf69}, 134416 (2004).
\bibitem{Katine} R. H. Koch, J. A. Katine, and J. Z. Sun, Phys. Rev. Lett. {\bf92}, 8 (2004).
\bibitem{APL} D. Pinna, Aditi Mitra, D. L. Stein, and A. D. Kent, arXiv:1205.6509 (2012).
\bibitem{Bedau} D. Bedau {\sl et al.}, Appl. Phys. Lett. {\bf97}, 262502 (2010).
\bibitem{Palacios} J. L. Garcia-Palacios and F. J. Lazaro, Phys. Rev. B {\bf58}, 22 (1998).
\bibitem{Taniguchi} T. Taniguchi and H. Imamura,Phys. Rev. B {\bf83}, 054432 (2011).
\bibitem{Taniguchi3} T. Taniguchi and H. Imamura,Appl. Phys. Express {\bf5}, 063009 (2012).
\bibitem{Karatsas} I. Karatsas and S. Shreve, {\em Brownian Motion and Stochastic Calculus, 2nd ed.}(Springer-Verlag, New York, 1997).
\bibitem{Nguyen} H. Nguyen, {\em GPU Gems 3} (Addison-Wesley Professional, 2007).
\bibitem{Ecuyer} P. L'Ecuyer, Operations Research {\bf44}, 5 (1996), pp. 816-822.
\bibitem{Press} W.H. Press, S.A. Teukolsky, W.T. Vetterling and B.P. Flannery, {\em Numerical recipes in C: The art of scientific computing} (Cambridge University Press, 1992).
\bibitem{Zwanzig} R. Zwanzig, {\em Nonequilibrium Statistical Mechanics} (Oxford Univerity Press, Oxford, UK, 2001).
\bibitem{Taniguchi2} T. Taniguchi and H. Imamura, Phys. Rev. B {\bf85}, 18440 (2012).
\bibitem{Kubo} R. Kubo, N. Hashitsume, Supp. Prog. Theor. Phys. {\bf 46}, pp. 210-220 (1970).
\bibitem{Rumelin} W. Rümelin, SIAM Journal on Numerical Analysis {\bf 19}, 3 (1982), pp. 604-613
\bibitem{Coffey} W. T. Coffey, Phys. Rev. B {\bf58}, 3249 (1998).
\bibitem{Mitra06} Aditi Mitra, So Takei, Yong Baek Kim, and A. J. Millis, Phys. Rev. Lett. {\bf 97}, 236808 (2006) 
 
%\bibitem{Vanden} R. V. Kohn, M. G. Reznikoff, and E. Vanden-Eijnden, J. Nonlinear Sci. {\bf15}: pp. 223–253 (2005).
%\bibitem{Serpico} G. Bertotti, I. Mayergoyz, and C. Serpico, {\em Nonlinear Magnetization Dynamics in Nanosystems} (Elsevier, Oxford, UK, 2009).

\end{thebibliography}
\end{document}